\documentclass[reprint,amsmath,amssymb,aps]{revtex4-2}

\usepackage{amsmath}
\usepackage{tipa}
\usepackage{bbm}
\usepackage{txfonts}
\usepackage{graphicx}
\usepackage{dcolumn}
\usepackage{bm}
\usepackage{amssymb}
\usepackage{latexsym}
\usepackage{color}
\usepackage{graphicx}
\usepackage{floatrow}
\usepackage{subfig}
\usepackage{caption}
\usepackage{ulem}
\usepackage[mathscr]{euscript}
\usepackage[colorlinks=true,linkcolor=blue,citecolor=blue,urlcolor=blue,]{hyperref}
\usepackage[doipre={doi:~}]{uri}

\begin{document}

\title{Anisotropy-induced Coulomb phase and quasiparticle zoo in the atomic monopole-spin hybrid system}
\author{Shao-Jun Li$^1$}
\author{Xiang Gao$^1$} 
\author{Xue-Ting Fang$^1$}
\author{Lushuai Cao$^{1}$}\email[E-mail: ]{lushuai$\_$cao@hust.edu.cn}
\author{Peter Schmelcher$^{3,4}$}
\author{Zhong-Kun Hu$^{1,2}$}\email[E-mail: ]{zkhu@hust.edu.cn}
\affiliation{$^1$MOE Key Laboratory of Fundamental Physical Quantities Measurement $\& $ Hubei Key Laboratory of Gravitation and Quantum Physics, PGMF and School of Physics, Huazhong University of Science and Technology, Wuhan 430074, People’s Republic of China \\
$^2$Wuhan Institute of Quantum Technology, Wuhan 430206, People’s Republic of China\\
$^3$Zentrum für optische Quantentechnologien, Universität Hamburg, Luruper Chaussee 149, 22761 Hamburg, Germany \\
$^4$The Hamburg Centre for Ultrafast Imaging, Universität Hamburg, Luruper Chaussee 149, 22761 Hamburg, Germany
}
\date{\today}

\begin{abstract}
Quantum simulation of a monopole-spin hybrid system is performed on basis of a dipolar ultracold gas in a ladder lattice. 
The site-occupation states of the dipolar ladder lattice gas can spontaneously emulate
both the monopole and spin excitations. The hopping of the atoms induces a
particle conversion process between spin and monopole pairs, and the dipole-dipole interaction
determines the spin-spin, spin-monopole and monopole-monopole interactions.
The anisotropic nature of the dipole-dipole interaction allows hereby for a  flexible engineering
of the designed hybrid system, and for a significant tunability of the interaction strengths.
 As a result, we encounter a rich phase diagram, and specifically a self-assembled Coulomb phase arises, 
in which monopoles and spins coexist and are orderly arranged according to the local Gauss's law.
The Coulomb phase hosts a zoo of different types of quasiparticles, and provides the possibility
to simulate various phenomena in particle physics, such as a degenerate vacuum, 
particle decay and conversion processes.
Our work provides a significant extension of the scope of quantum simulations based on the anisotropy of dipolar interactions.
\end{abstract}


\maketitle

% I:Introduction
\section{Introduction}
The site-occupation degree of freedom (SOD) has become an important resource for ultracold atoms 
confined in optical lattices, with the advantages of a high controllability,
low decoherence due to coupling to the environment and the direct detection possibility by the quantum gas microscope. 
In quantum simulation with ultracold atoms, the site-occupation states have been adapted to 
simulate different targets, ranging from spin chains 
 in condensed matter physics \cite{Sachdev2002,Simon2011,Meinert2013,Buyskikh2019,Cao2015,
 Ketterle2016,Ketterle2017,Lee2017,Liao2021,Liao2022,Gao2022,Gao2023} to
the gauge fields in particle physics \cite{Wiese2013,Pan2020,Pan2022,Paulson2021,Halimeh2022,Zhai2022,Zhai2023}.
Additional potentials are explored to manipulate the SOD and to engineer the underlying effective Hamiltonian, 
such as the case of  the linear tilt potential \cite{Sachdev2002,Simon2011,Meinert2013,Buyskikh2019} as well as 
staggered and longer-period lattices \cite{Cao2015,Ketterle2016,Ketterle2017,Gao2022,Gao2023,Pan2020,Pan2022,Halimeh2022,Zhai2022,Zhai2023}.
Such Hamiltonian engineering, however, normally bears the limitition that 
the simulation is restricted to a narrow parameter space and/or Hilbert space of the target system, 
and it is highly desirable to introduce more flexible tools for
engineering the Hamiltonian and to extend the scope of quantum simulation with the SOD.

The dipole-dipole interaction (DDI) provides us with a high controllability of the SOD,
due to its long interaction range and anisotropic nature. 
The DDI can be induced by polar molecules, Rydberg atoms, atoms with light-induced
dipole moments and atoms with a magnetic dipole moment \cite{Lahaye2009,Chomaz2022}.
The anisotropy of DDI can exert a strong influence on the static and dynamical properties of ultracold
atoms, such as modifying their spatial \cite{Yi2000,Martik2001,Petrov2012,Baillie2014,Zhang2021} and momentum 
distribution \cite{Stefano2003,Stuhler2005,Aikawa2014,Zhang2020} 
as well as the phonon excitation spectrum \cite{Bismut2012}  and nonlinear solitons \cite{Pedri2005,Tikhonenkov2008} .
It is then of direct interest to explore the DDI, particularly its anisotropy nature, for
the engineering of the effective Hamiltonian, 
which holds promises to broaden the scope of quantum simulation employing the SOD. 

In this work, we explore dipolar ultracold atoms confined in a ladder lattice, i.e. the dipolar
ladder gas (DLG), for the quantum simulation of the interaction between magnetic monopoles and spins.
In this simulation scheme, the monopole and spin are both mapped to the SOD,
which allows accounting for particle conversion processes between monopoles
and spins by the atomic hopping in the lattice.
Besides, the DDI is employed to engineer the spin-spin,
spin-monopole and monopole-monopole interactions and the excitation energy of the monopole.
It turns out that the anisotropic nature of the DDI, characterized by the relative angle
between the dipole moment of the atoms and the orientation of the ladder lattice, provides a flexible
tunning of the coupling strengths and the excitation energies, which leads to a rich phase diagram 
evolving from the spin to the charge sectors in the Hilbert space as the direction of the dipole moment is
varied. Particularly, we identify a new type of Coulomb phase as part of the phase diagram,
with the monopoles and spins spatially arranged according to Gauss's law.
Comparing to those appearing in frustrated spin systems
 \cite{Huse2003,Henley2010,Perrin2016,Alan2019,Ran2023}
and atomic simulators for gauge fields
\cite{Pan2020,Pan2022,Paulson2021,Halimeh2022,Zhai2022,Zhai2023}, the Coulomb phase proposed in this work
presents its own uniqueness that, for one thing monopoles and
spins are self-assembled into the spatially ordered arrangement, instead of induced by
external potentials, and for another
this Coulomb phase also hosts a zoo of different quasiparticles, such as bounded monopole pairs and
spin clusters. These quasiparticles  present behaviors of particle conversion and decay, which
extend our simulation scope to the particle physics using the SOD.

Our work is organized as follows. In Sec. \ref{setup} we introduce the setup the underlying Hamiltonian
and the effect of the anisotropic DDI. In Sec. \ref{transition} we present the phase diagram of the ladder lattice. Section \ref{zoo} 
contains an investigation of the zoo of quasiparticles in the coulomb phase. A brief summary and discussion
 are proviced in Sec. \ref{summary}.

\section{Setup, Hamitonian and anisotropy}\label{setup}
\subsection{Simulation scheme}
\begin{figure*}[htbp]
\sidesubfloat[]{\includegraphics[trim=70 130 70 10,width=0.9\textwidth]{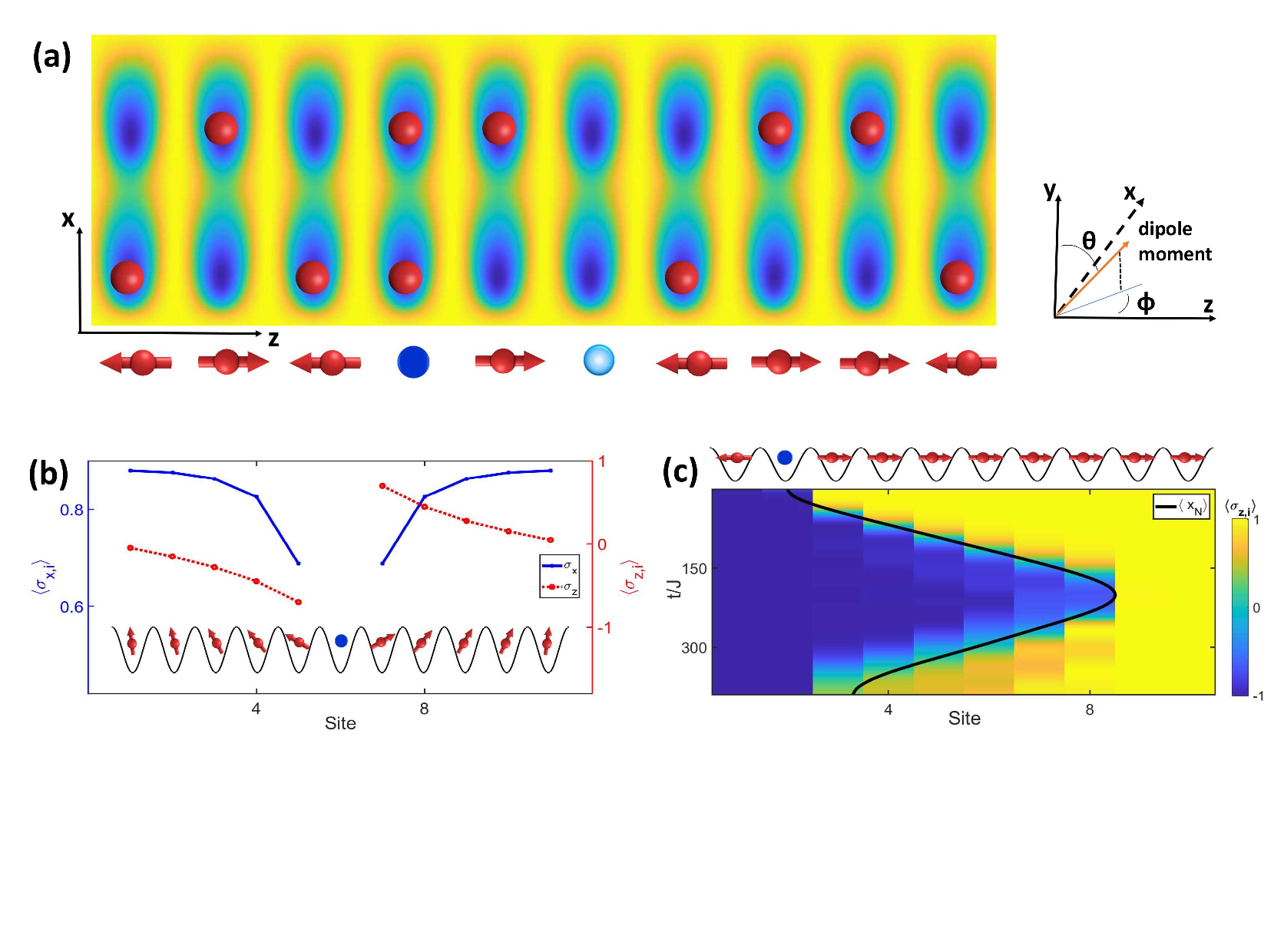} }
\caption{(color online).  (a) The generalized pseudospin mapping, with the upper and the lower panel
referring to the original DLG system and the simulated monopole-spin hybrid system, respectively.
(b) The magnetization $\langle \sigma_x^i\rangle$ (blue solid line) and $\langle \sigma_z^i\rangle$
(red dotted line) of the spins around a doublon defect, and the inset at the bottom shows the spin polarization
of these spins according to the calculated expectation values. The parameters
for the calculation are $\left(J,J_1,V_{\rm{dd}}/a^3,\gamma,\theta,\phi\right)=\left(1,0.1,20,48^\circ,35^\circ,17^\circ\right)$  
(c) The dynamical evolution of $\langle \sigma_z^i\rangle$ during the hopping of the doublon. The upper
 pannel shows the initial state, and in the main figure the solid line indicates the mean location of
 the doublon, i.e.  $\langle \hat{D}_i\rangle$. The color quantifies the polarization.
The parameters
for the calculation are $\left(J,J_1,V_{\rm{dd}}/a^3,\gamma,\theta,\phi\right)=\left(1,0.6,80,48^\circ,50^\circ,30^\circ\right)$}
\label{ladder}
\end{figure*}

We explore the DLG system, comprised of spin-polarized fermionic atoms confined in 
a ladder lattice, for the quantum simulation of a monopole-spin hybrid system.
The ladder lattice is composed of two strongly coupled 
one-dimensional lattices aligned parallel, as the two legs of the ladder lattice. 
Each rung of the lattice contains two sites and forms a supercell of the lattice.
The ultracold atoms can hop between nearest neighbor sites along each leg and within 
the rung, and interact with each other through the dipole-dipole interaction, 
with the dipole moment of all atoms along the same direction.
As shown in Fig. \ref{ladder}(a), the orientation of the dipole moment with respect to the
ladder lattice can be specified by the azimuthal and polar angles $\left(\theta,\phi\right)$, 
which determines the anisotropic nature of the DDI.
Under the tight-binding approximation, the Fermi-Hubbard Hamiltonian of the $N$-rung lattices given as:
\begin{equation}
\begin{aligned}
{\hat H_{\rm{HB}}} &= - J\sum\limits_{i = 1}^N \left( {{\hat f}_{i,U}^\dag {{\hat f}_{i,D}} + {\rm{H}}.{\rm{c}}.} \right) -J_1\sum\limits_{i=1}^N\sum\limits_{s=U,D}\left(  {\hat f}_{i,s}^\dag{\hat f}_{i+1,s}+{\rm{H.c.}} \right) \\  
& +\frac{1}{2} \sum\limits_{\langle\left( i,s_1\right)\neq\left(j,s_2\right)\rangle}{U_{i,s_1,j,s_2} }\hat n_{i,s_1} \hat n_{j,s_2}.\label{H_HB}
\end{aligned}
\end{equation}
In the above equation, $\hat f_{i,s}^\dag\left(\hat f_{i,s}\right)$ creates (annihilates) an atom
in the $i$-th rung and $s$-leg, and the site occupation operator is defined as
$\hat n_{i,s_1}\equiv \hat f_{i,s_1}^\dag \hat f_{i,s_1}$, with $s=U/D$ denoting the upper/lower legs.
The first two terms of $\hat H_{\rm{HB}}$ 
denote the intra- and inter-cell hopping, respectively, with the condition of $J\gg J_1$. 
The last term of $\hat H_{\rm{HB}}$ denotes the dipole-dipole interaction between atoms in different sites,
and the anisotropic effect of ${U_{i,s_1,j,s_2} }$ is the main concern of this work, of which the
explicit dependence on $\left(\theta,\phi\right)$ and the consequences will be addressed in the following subsection. 
In this work, we assume a moderate interaction strength, which guarantees truncating
the dipole-dipole interaction to nearest-neighbor (NN) interaction.

The simulation of the spin-monopole hybrid system is based on the pseudospin mapping, 
which maps the site-occupation states to the spin and monopole excitations.
Each supercell of the ladder lattice presents four occupation states, and in the $i$-th cell,
for instance, the four occupation states are 
$\{|1,0\rangle_i,|0,1\rangle_i,|1,1\rangle_i,|0,0\rangle_i\}$,
where $|n_1,n_2\rangle_i$ denotes the upper and lower site of the $i$-th cell occupied by 
$n_1$ and $n_2$ atoms, repsectivley.
The standard pseudospin mapping transfers the single-occupation state $\left|0,1\rangle_i\right.$ 
($\left|0,1\rangle_i\right.$) to the spin state $\left| \leftarrow\rangle_i \right.$ 
($\left| \rightarrow\rangle_i\right.$). The doublon $\left|1,1\rangle_i\right.$
and holon $\left|0,0\rangle_i\right.$ correspondingly manifest themselves as magnetic impurities
located at the $i$-th cell.
We refer to the spin and charge sector as the Hilbert subspace spanned by the
basis states composed of only spins and impurites, respectively. Besides, 
the basis states composed of both spin and impurities span the mixed sector. The generalized
pseudospin mapping is shown in Fig. \ref{ladder}(a), with the upper and bottom panel indicating
the original DLG system and the simulated impurity-spin hybrid system, respectively.

Under the pseudospin mapping, ${\hat H_{\rm{HB}}}$ is transformed to:
\begin{equation}
\begin{aligned}
\hat H_{\rm{eff}}&=\hat H_{\rm{spin}}+\hat H_{\rm{mnp}}+\hat H_{\rm{spin-mnp}}+\hat H_{\rm{conv}},\label{Heff} \\
\end{aligned}
\end{equation}
\begin{equation}
\begin{aligned}
\hat H_{\rm{spin}}&=E_{\rm{0}}+\sum\limits_{i}^N{\left(V_{\rm{SS}}{\hat \sigma}_z^i{\hat \sigma}_z^{i+1}-J{\hat \sigma}_x^{i}\right)} \notag\\
\end{aligned}
\end{equation}
\begin{equation}
\begin{aligned}
\hat H_{\rm{mnp}}&=\sum\limits_{i=1}^N{V_{\rm{MM}}\left(\hat D_i-\hat H_i\right)\left(\hat D_{i+1}-\hat H_{i+1}\right)}\\
                 &+\sum\limits_{i=1}^N{\left(E_{\rm{N}}\hat D_i+E_{\rm{S}}\hat H_i\right)}, \notag
\end{aligned}
\end{equation}
\begin{equation}
\begin{aligned}
\hat H_{\rm{spin-mnp}}&=\sum\limits_{i=1}^N {V_{\rm{MS}}\left(\hat D_i-\hat H_i\right) \left({\hat \sigma}_z^{i+1}-{\hat \sigma}_z^{i-1}\right)} \\&-J_1\sum\limits_{i;s=\rightarrow,\leftarrow} \left(\hat s_{s,H}^i \hat s_{H,s}^{i+1}+\hat s_{D,s}^i \hat s_{s,D}^{i+1}+H.c.\right), \notag\\
\end{aligned}
\end{equation}
\begin{equation}
\begin{aligned}
\hat H_{\rm{conv}}&=-J_1\sum\limits_{i} \left(\hat s_{D,\rightarrow}^i\hat s_{H,\leftarrow}^{i+1} -\hat s_{D,\leftarrow}^i\hat s_{H,\rightarrow}^{i+1}-\hat s_{\rightarrow,H}^i\hat s_{\leftarrow,D}^{i+1}\right.  \\
&\left. +\hat s_{\leftarrow,H}^i\hat s_{\rightarrow,D}^{i+1}+H.c. \right).\notag 
\end{aligned}
\end{equation}
In the above equations, ${\hat \sigma}_x^i\equiv {\hat f}_{i,1}^\dag {\hat f}_{i,2}+\rm{H.c.}$
and ${\hat \sigma}_z^i\equiv \hat n_{i,1}-\hat n_{i,2}$ are the Pauli matrices acting on the
spin at the $i$-th cell.
$\hat D_i\equiv \hat n_{i,1}\hat n_{i,2}$ and 
$\hat H_i\equiv\left(1-\hat n_{i,1}\right) \left(1-\hat n_{i,2}\right)$ are defined as the site-occupation
operator of the doublon and holon at the $i$-th cell, respectively.
$\hat s_{s_1,s_2}^i\equiv\left|s_1\rangle_i \langle s_2\right|_i$ refers to the particle conversion
operator, with $s_1,s_2 \in \left\{ \leftarrow,\rightarrow,D,H\right\}$.

$\hat H_{\rm{spin}}$ in $\hat H_{\rm{eff}}$ indicates that the pesudospins are exposed to an 
effective transverse magnetic
field, and interact with each other via the Ising type spin-spin interaction,
 thereby forming a transverse Ising spin chain. 
$\hat H_{\rm{mnp}}$ defines the excitation energy and the 
interaction between the impurities, which demonstrates a counterintuitive effect that 
a holon can exert a repulsive (attractive) interaction to the holon (doublon) in its neighbor
cell, even though the cell of holon contains no atoms. 
This counterintuitive effect is due to the fact that the zero-energy point is chosen as 
the ground state energy of a lattice fully filled with pseudospins, and the interaction
between two holons, for instance, simply describes that the two holons dopped to 
the pseudospin chains prefer to be separated to lower the total energy of the system.

The coupling between the spin and the 
doublon (holon) is given in $\hat H_{\rm{spin-mnp}}$, and the first term
refers to the interaction between an impurity and its neighbor spins, which
mimics an effective magnetic field exerted on the spins by the impurity.
The effective magnetic field of the doublon (holon) polarizes the spins on both sides away from (towards to)
the corresponding impurity, and resembles the singular magnetic field around the north (south) monopoles.
We will demonstrate that this indeed fulfills the fingerprint of the magnetic monopole.
The second term of $\hat H_{\rm{spin-mnp}}$ manifests as the exchange interaction between the
doublon (holon) and its neighbor pseudospin. 
The last term of $\hat H_{\rm{eff}}$, i.e. $\hat H_{\rm{conv}}$, describes the particle conversion
process between a pair of pseudospins and a doublon-holon pair.

We demonstrate that the doublon and holon can simulate the north and south monopole, respectively,
which should fullfill the two fingerprints of the magnetic monopoles, including the singular magnetic
field and the Dirac string \cite{Savage2003,Pietil2009,Pietil2009_1,Ruokokoski2011,Ray2014,Ray2015,
Tiurev2016,LiJi2017,Ollikaninen2017,Sugawa2018,Ollikainen2019,
Tiurev2019,Mithun2022,Henley2010,Perrin2016,Alan2019}.
The singular magnetic field is witnessed by the spin polarization around monopoles,
and the Dirac string can be modeled by the dynamical effect that the hopping of the
monopole flips the spins along the hopping path, such as in the spin-ice 
\cite{Henley2010,Perrin2016,Alan2019} and double-well superlattice quantum gas schemes \cite{Gao2022} .
In order to confirm the singular magnetic field effect, 
Fig. \ref{ladder}(b) presents the spin polarization in terms of $\langle \sigma_z^i\rangle$ and 
$\langle \sigma_x^i\rangle$ of the spins around a doublon fixed to the middle cell of the ladder lattice,
of which $\langle ...\rangle$ denotes the expectation with respect to the ground state.
It can be found that the neighboring spins on both sides of the doublon
are polarized away from the impurity, and indeed mimic the singular magnetic field of a north monopole.
The doublon also presents the second fingerprint of the Dirac string effect,
and Fig. \ref{ladder}(c) shows the corresponding dynamical evolution of the DLG system, 
of which the initial state is taken as doping a doublon to the second left cell of the 
lattice and relaxing the remaining spins to the ground state. 
It can be observed that initially all the spins are polarized according to the
singular magnetic field, and as the doublon hops forth and back in the lattice, 
indicated by $\langle \hat{D}_i\rangle$ (solid dark line),
the spins are indeed flipped along the hopping path as done by the Dirac string effect.
Figures \ref{ladder}(b) and (c) demonstrate that the doublon carries the
two fingerprints and can simulate the north monopole. Similarly,
the south monopole can be simulated by the holon.
In the following we will refer to the doublon and holon as
the north and south monopole, respectively.

\subsection{Anisotropy effect}
The anisotropy effect of the dipole-dipole interaction refers to the dependence of the interaction strength
on the angle between the direction of the relative displacement of the two dipoles 
and the dipole moment, as:
\begin{equation}
\begin{aligned}
    U_{i,s_1,j,s_2}=V_{\rm{dd}}\left(1-3\cos^2\chi_{i,s_1,j,s_2}\right)/{R^3_{i,s_1,j,s_2}},
\end{aligned}
\end{equation}
with the coupling constant $V_{\rm{dd}}$ \cite{Lahaye2009},
the relative distance $R_{i,s_1,j,s_2}$, and the relative angle $\chi_{i,s_1,j,s_2}$ between the 
dipole moment and the relative displacement. The anisotropy effect  implies that
the dipole-dipole interaction is different along the leg and the rung of the ladder lattice. 

Truncated to nearest neighbor, the dipole-dipole interaction reads:
\begin{equation}
\begin{aligned}
\hat U_{\rm{NN}} &= \sum\limits_{i=1}^N V_{\rm{0}}\left( \theta,\phi \right) \hat n_{i,1}\hat n_{i,2} 
                  +\sum\limits_{i=1}^N V_{\rm{para}}\left( \theta,\phi\right) \hat n_{i,1}\hat n_{i+1,1}  \\
                 &+\sum\limits_{i=1}^N V_{\rm{para}}\left( \theta,\phi \right) \hat n_{i,2}\hat n_{i+1,2} 
                  +\sum\limits_{i=1}^{N} V_{\rm{UD}} \left( \theta,\phi\right) \hat n_{i,1}\hat n_{i+1,2} \\
                 &+\sum\limits_{i=1}^N V_{\rm{DU}}\left( \theta,\phi \right) \hat n_{i,2}\hat n_{i+1,1},\label{HV}\\
\end{aligned}
\end{equation}
with $(\theta,\phi)$ denoting the azimuthal and polar angles of the dipole moment, as shown in Fig.\ref{ladder}(a).
In $\hat U_{\rm{NN}}$, the interaction is truncated to that between nearest neighbors, and the on-site interaction term is
excluded as fermionic atoms are considered here. 
$V_{\rm{0}}$ denotes the interaction strength of atoms in the same cell, and 
$V_{\rm{para}}$, $V_{\rm{UD}}$ and $V_{\rm{DU}}$ refer to the interaction strength between two nearest neighbors, with
the atoms in the same and different legs, respectively. 
The interaction strengths can be derived as:
\begin{subequations}
\begin{align}
V_{\rm{0}}\left( \theta,\phi \right) &=\frac{V_{\rm{dd}}\left( 1-3\sin^2 \theta \sin^2 \phi \right)}{a^3 \tan^3\gamma},\label{Vcell}
\\V_{\rm{para}}\left( \theta,\phi\right)  &=\frac{V_{\rm{dd}}\left( 1-3\sin^2\theta \cos^2\phi \right)}{a^3},\label{V00}  \\
V_{\rm{UD}}\left(\theta,\phi\right) &=\frac{V_{\rm{dd}}\cos^3 \gamma\left[ 1-3\sin^2\theta \cos^2\left( \phi+\gamma\right)\right]}{a^3},\label{V01} \\
V_{\rm{DU}}\left(\theta,\phi\right) &=\frac{V_{\rm{dd}}\cos^3 \gamma\left[ 1-3\sin^2\theta \cos^2\left( \phi-\gamma\right)\right]}{a^3},\label{V10}
\end{align}
\end{subequations}
where $\gamma=\arctan{\left( b/a\right)}$, with $a$ ($b$) denoting the length of the rung (leg).

\begin{figure}[htbp]
\includegraphics[trim=50 80 70 5,width=0.95\textwidth]{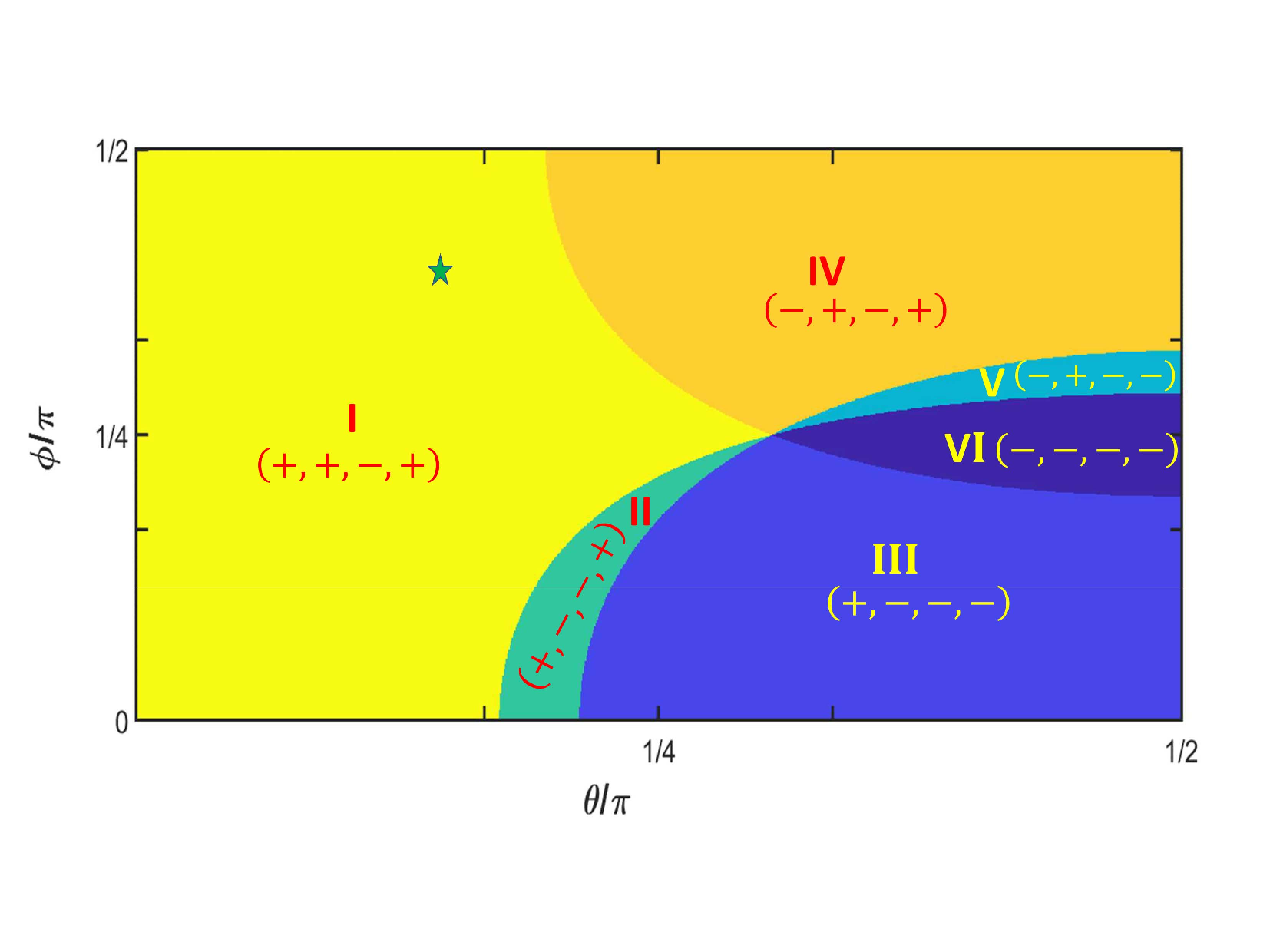}
\caption{(color online). Anisotropy effect of the dipole-dipole interaction in terms of the sign
of $\left(E_{\rm{NS}},V_{\rm{SS}},V_{\rm{MS}},V_{\rm{MM}}\right)$ as a function of 
$\left(\theta,\phi\right)$ for $\gamma=48^\circ$. The different colored regimes mark the parameter regimes with 
different signs of $\left(E_{\rm{NS}},V_{\rm{SS}},V_{\rm{MS}},V_{\rm{MM}}\right)$.
 }
\label{anisotropy}
\end{figure}

Under the pseudospin mapping, the coefficients of the effective Hamiltonian, including the strengths of the spin-spin interaction $V_{\rm{SS}}$, the monopole-spin interaction $V_{\rm{MS}}$ and
the monopole-monopole interaction $V_{\rm{MM}}$, as well as the total excitation energy $E_{\rm{NS}}=E_{\rm{N}}+E_{\rm{S}}$ of a north
monopole $E_N$ and a monopole $E_S$, where $E_{\rm{N}}\left(E_{\rm{S}}\right)$ is the excitation energy of the north(south) monopole, are dependent on the DDI, and read:
\begin{subequations}
\begin{align}
V_{\rm{SS}}=&\frac{V_{\rm{dd}}\left[ 1-3\sin^2\theta \cos^2\phi-\cos^3 \gamma\left(1-\alpha/2\right) \right]}{2a^3} \\   
V_{\rm{MS}}=&\frac{-3V_{\rm{dd}}\cos^3\gamma\sin^2 \theta \sin2\phi\sin2\gamma }{4a^3},\\
V_{\rm{MM}}=&\frac{V_{\rm{dd}}\left[ 1-3\sin^2\theta \cos^2\phi+\cos^3 \gamma\left(1-\alpha/2\right) \right]}{2a^3} \\ 
E_{\rm{NS}}=&\frac{V_{\rm{dd}}\left( 1-3\sin^2 \theta \sin^2 \phi \right)}{a^3 \tan^3\gamma}.
\end{align}
\end{subequations}
where $\alpha=3\sin^2\theta\left(1+\cos2\phi\cos2\gamma\right)$.

Figure 2 shows the sign dependence of $(E_{\rm{NS}},V_{\rm{SS}},V_{\rm{MS}},V_{\rm{MM}})$
on $(\theta,\phi)$, which highlights the tunability of the target Hamiltonian of the
 monopole-spin hybrid system with the anisotropy of the DDI. Figure \ref{anisotropy} is divided into different regimes according to the sign of the 
considered parameters, of which the excitation energy of the monopoles $E_{\rm{NS}}$ is positive in regimes 
I-III but switches to negative values in the other regimes. This suggests that the monopole-spin hybrid system
would undergo a phase transition from the spin sector to the charge sector as $(\theta,\phi)$
evolves from regimes I-III to IV-VI.
Moreover, within the charge and the spin sectors, the sign of the corresponding
interactions such as $(V_{\rm{SS}},V_{\rm{MS}},V_{\rm{MM}})$ also changes with $(\theta,\phi)$,
which indicates rich phase transitions within these regimes. The expected phase transitions indicate that the ladder lattice loaded with dipolar atoms can simulate a range of 
phenomena and effects in e.g. particle physics.

\section{Phase diagram}\label{transition}
\begin{figure}[htbp]
\includegraphics[trim=70 80 60 10,width=0.95\textwidth]{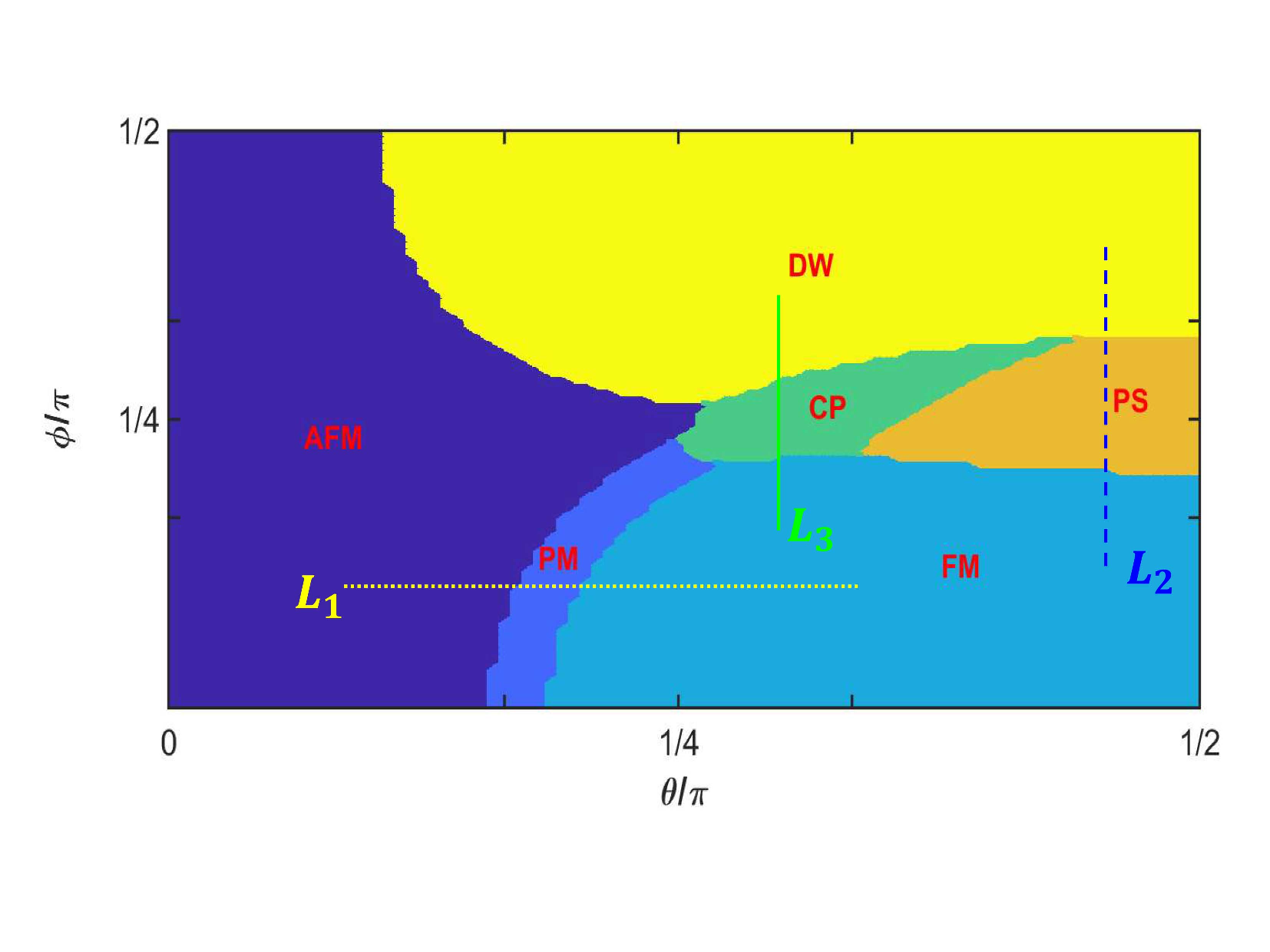}
\caption{(color online). The phase diagram of the atomic monopole-spin hybrid system simulated by the dipolar
ladder lattice gas. The simulation is performed on the twelve-cell ladder lattice loaded with $N=12$ fermions,
with $J=1,J_1=0.1,V_{\rm{dd}}/a^3=20,\gamma=48^\circ$. }
\label{phase}
\end{figure}

In this section, we present the numerical results on the phase diagram of the hybrid monopole-spin
system, with the simulation performed on an $12$-cell ladder lattice with periodic boundary 
conditions and unit filling per cell, i.e. $N=12$ fermions loaded in the lattice.
The numerical results are obtained by the exact diagonalization with $\hat H_{HB}$,
in order to avoid any artifact effect brought by $\hat H_{\rm{eff}}$.
Figure \ref{phase} shows the phase diagram with respect to $(\theta,\phi)$, and
before diving into the detailed analysis of different phases, let us take a general
look at the phase diagram: First of all, each colored regime of Fig. \ref{anisotropy}
is dominated by a particular phase, indicating that the quantum phase transition is driven by the sign
switching of $\left(E_{\rm{NS}},V_{\rm{SS}},V_{\rm{MS}},V_{\rm{MM}}\right)$. 
For instance,  the antiferromagnetic (AFM),
paramagnetic (PM) and ferromagnetic (FM) phases matches well with the regimes I-III, 
respectively, where the excitation energy of monopoles $E_{NS}$ are positive. In regimes IV-VI, with
$E_{NS}<0$, the ground state evolves to the charge sector, and the
charge density wave (DW) and phase separation (PS) states composed of monopoles arise.
More importantly, a phase appears around the intersection point of these sign-attributed regimes,
and turns out to be the Coulomb phase (CP), in which the monopoles and spins coexists and are
arranged according to the local Gauss' law. 
CP proposed in this work possesses the uniqueness that the spatial ordered arrangement is self-assembled
by the interactions between the monopoles and spins, instead of through external potential engineering, and
more importantly, this CP hosts a zoo of quasiparticles, which can be explored for simulating various 
particle-physics phenomena and extend the scope of the quantum simulation with ultracold lattice atoms.
It is also worth mentioning that the DLG simulator proposed in this work is within the reach of
current experimental platforms, such as the Hubbard quantum simulator composed of lattice erbium atoms \cite{Greiner2023}, which meets the major requirements, 
including the interaction strength and the stability control of the direction of the dipole moment .

\subsection{\label{spin phase transition}Phase transition in the spin sector}
We first inspect the phases lying in the spin sector, which occupies the parameter regimes I, II and III 
in Fig. \ref{anisotropy}. Since in these regimes $E_{\rm{NS}}$ is positive and prevents the pair excitation 
of monopoles, the ground state resides in the spin sector, and the dominant parameter 
of the phase transition is then the spin-spin
interaction strength $V_{\rm{SS}}$. $V_{\rm{SS}}$ switches sign from positive to negative as $(\theta,\phi)$ moves
from regime I to III and is expected to drive a phase transition between AFM to FM phases. 
In the intermediate regime of $V_{\rm{SS}}\approx0$, moreover, the transverse magnetic field with strength $J$
dominates over $V_{\rm{SS}}$ and prefers all the spins along its direction, which should result
in a PM phase in the intermediate regime.

\begin{figure}[htbp]
\includegraphics[trim=0 60 0 20,width=1.3\textwidth]{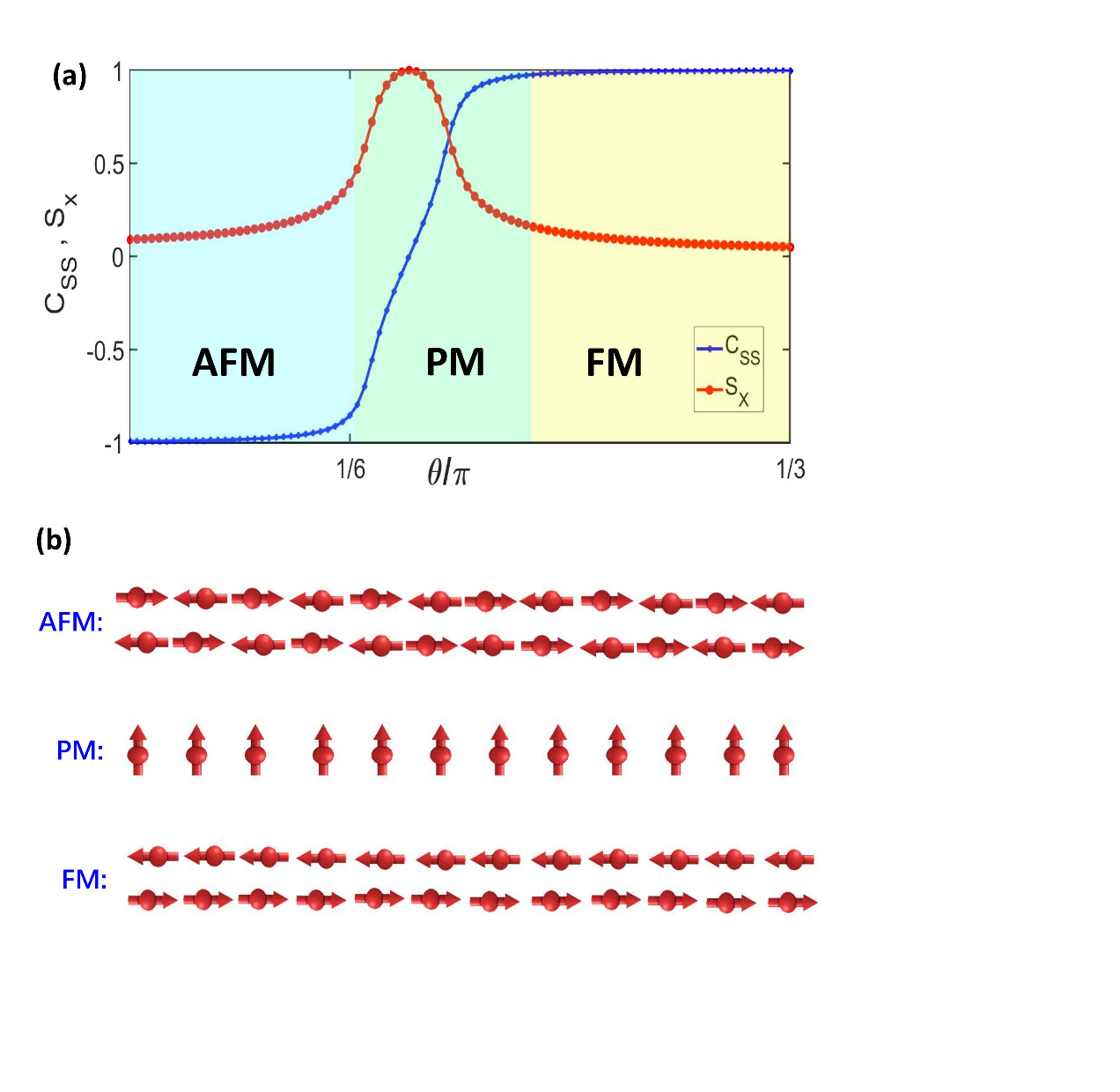}
\caption{(color online).  (a) $C_{\rm{SS}}$ and $S_{\rm{x}}$ along the cutting (dotted yellow) line $\rm{L}_1$ of fig.  \ref{phase}. (b) sketch of AFM,PM and FM states.}
\label{Css}
\end{figure}

In order to verify the above anticipation of the FM-PM-AFM transition in the spin sector,
we determine the evolution of the spin-spin correlation $C_{\rm{SS}}$ and
the magnetization $S_{\rm{x}}$ along the cutting line $L_1$ acrossing the
three regimes in the phase diagram (dotted yellow line in Fig. \ref{phase}),
which are defined as:
\begin{equation}
\begin{aligned}
C_{\rm{SS}}=\langle \sum\limits_{i=1}^N{\hat \sigma_z^i\hat \sigma_z^{i+1}}\rangle/N,\\
S_{\rm{x}}=\langle \sum\limits_{i=1}^N{\hat \sigma_x^i}\rangle/N,\label{Sx}
\end{aligned}
\end{equation}
where $\langle...\rangle$ denotes the expectation with respect to the ground state.
In Fig. \ref{Css}, $C_{\rm{SS}}$ shows two regions: in the first (second) region 
$C_{\rm{SS}}$ approaches $-1$ ($1$) with $S_{\rm{x}}$ vanishing, 
which is a direct signature that the system resides in a  AFM (FM). 
Phase in the intermediate regime between the AFM and FM, 
$S_{\rm{x}}$ shows a relatively broad peak, and this indicates that all spins are aligned along the x-axis,
i.e. the direction of the transverse magnetic field, which results in the PM in the narrow regime between the two regions of AFM and FM. 
The spin configurations belonging to the three magnetic phases are illustrated in Fig. \ref{Css} (b).
It is worth noticing that a similar phase transition to AFM-PM-FM has also been simulated with dipolar BEC loaded into a ladder lattice \cite{Lee2017}, which actually simulates the classical Ising spin chain.

\subsection{\label{pair excitation} Phase transition in the charge sector}
We proceed to the parameter regimes IV, V and VI in Fig. \ref{anisotropy}, which
share the common property of $E_{\rm{NS}}<0$, while $V_{\rm{MM}}$ switches sign among these regimes.
The negative $E_{\rm{NS}}$ implies that the ground state shifts to the charge sector in these regimes,
since the excitation of monopoles lowers the total energy. The
$V_{\rm{MM}}$ in the charge sector drives a phase transition between 
DW and PS phases. 

\begin{figure}[htbp]
\includegraphics[trim=0 50 0 10,width=1.3\textwidth]{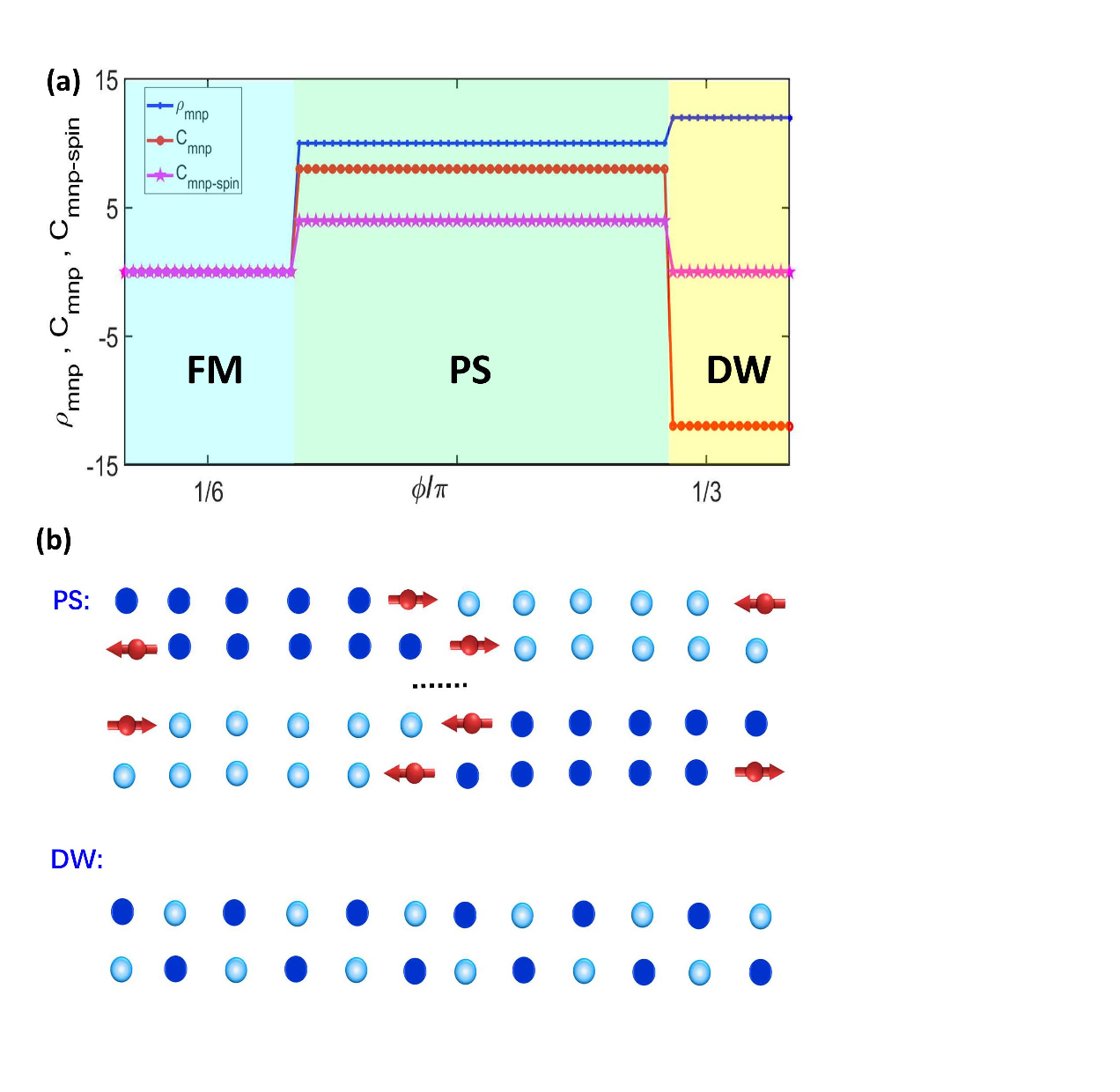}
\caption{(color online).  (a) $\rho_{\rm{mnp}}$, $C_{\rm{mnp}}$ and $C_{\rm{mnp-spin}}$ along $L_2$ 
(dashed blue line) of fig. \ref{phase}.  (b) sketch of PS and DW states.}
\label{correlations}
\end{figure}

The DW-PS phase transition in the charge sector can be confirmed by 
the monopole density $\rho_{\rm{mnp}}$, the nearest neighbor monopole correlation $C_{\rm{mnp}}$
and the monopole-spin correlation $C_{\rm{mnp-spin}}$, which are defined as:
\begin{equation}
\begin{aligned}
\rho_{\rm{mnp}}&=\langle\sum\limits_{i=1}^N{\left(\hat D_i+\hat H_{i}\right)}\rangle,\\
C_{\rm{mnp}}&=\langle\sum\limits_{i=1}^N{\left(\hat D_i-\hat H_{i}\right)\left(\hat D_{i+1}-\hat H_{i+1}\right)}\rangle,\\
C_{\rm{mnp-spin}}&=\langle \sum\limits_{i=1}^N{\left(\hat D_i-\hat H_{i}\right)\left(\hat \sigma_z^{i+1}-\hat \sigma_z^{i-1}\right)}\rangle.
\end{aligned}
\end{equation}
Figure \ref{correlations} (a)
presents the evolution of $\rho_{\rm{mnp}}$, $C_{\rm{mnp}}$ and $C_{\rm{mnp-spin}}$
along the $\rm{L}_2$ line (dashed blue line) in the phase diagram, which demonstrates 
the transition between the FM, PS and DW phases.
As the system evolves from the FM to the PS phase, $\rho_{\rm{mnp}}$ increases from
nearly vanishing to a finite value, indicating that spins are converted to monopoles.
Moreover, $C_{\rm{mnp}}$ is positve in the PS phase, which indicates that monopoles of the same type are 
grouped together, giving rise to the phase separation state composed of a north- and south-monopole domain.
The positive $C_{\rm{mnp-spin}}$ further indicates two spins located at the boundaries of the two domains and
both pointing from the north- to the south-monopole domain. 
In the DW phase, as interaction between the north and south monopoles become attractive, the two monopole
domains break and the north and south monopoles are arranged in an alternating way to form the density wave states.
The twelve-fold degenerate PS and two-fold degenerate DW states are also sketched in 
Fig. \ref{correlations} (b).

It is also interesting to notice that the phase boundary between DW and AFM in Fig. \ref{phase} does quantitatively
not coincide with that between regime I and IV in Fig. \ref{anisotropy}, despite that the phase
transition between DW and AFM is mainly driven by the sign change of $E_{\rm{NS}}$. 
It can be found that the DW phase composed of monopoles extends to the parameter 
regime with $E_{\rm{NS}}>0$.
This quantitative mismatch between the two boundaries reflects the competition
among $E_{\rm{NS}}$, $V_{\rm{MM}}$ and $V_{\rm{SS}}$. For instance,
around the star mark in Fig. \ref{anisotropy} with $E_{\rm{NS}}>0$, 
which prefers the system residing in the spin sector, the ground state turns
out to be the DW in the charge sector, and this is attributed to the effect that
at the star mark, $V_{\rm{MM}}$ is much stronger than $V_{\rm{SS}}$, and the energy cost to convert a pair of spin into two monopoles is 
compensated by the interaction energy difference of $V_{\rm{MM}}-V_{\rm{SS}}$,
which results in the extension of the DW phase into regime I of Fig. \ref{anisotropy}.

\subsection{Coulomb phase}
At the boundaries of certain regimes in Fig. \ref{anisotropy}, 
there arises a novel phase, which we term Coulomb phase (CP), which is 
composed of spatially ordered monopoles and spins. 
The CP has been encountered in different setups, such as spin-monopole hybrid system,
realized by frustrated spin models \cite{Huse2003,Henley2010,Perrin2016,Alan2019,Ran2023},
and in lattice gauge field models composed of photons and charges 
\cite{Pan2020,Pan2022,Paulson2021,Halimeh2022,Zhai2022,Zhai2023}. In these systems,
CP is commonly characterized by Gauss's law, which regularizes the polarization of photons
(spins) around charges (monopoles). A similar Gauss' law can also be introduced for
 the CP in the DLG simulator, which can be further explored for the simulation of various
particle-physics phenomena, such as the vacuum degeneracy and particle decay/conversion processes.

\begin{figure}[htbp]
\includegraphics[trim=0 150 0 10,width=1.3\textwidth]{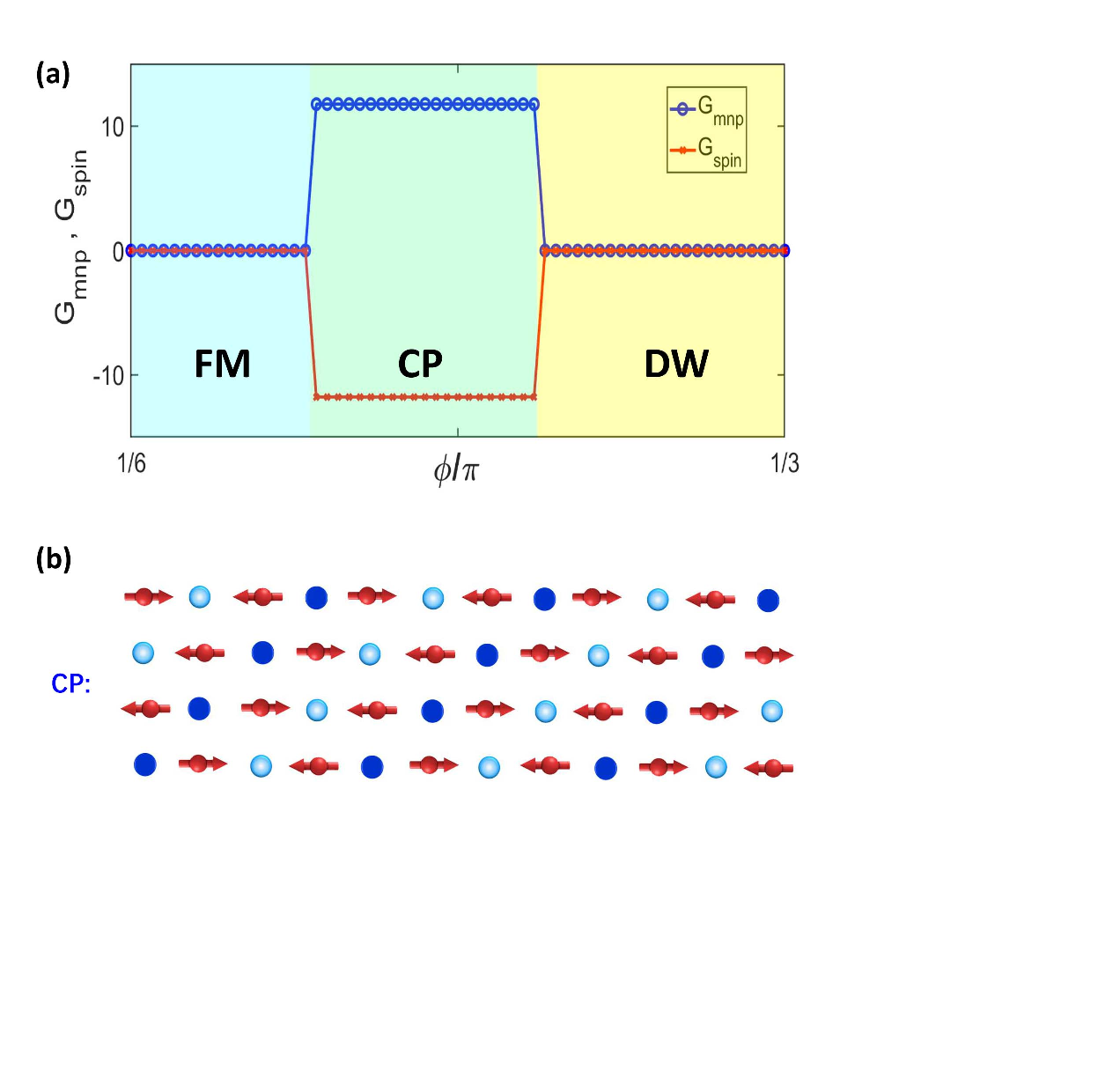}
\caption{(color online).  (a) Expectation values of $\hat G_{\rm{mnp}}$ and $\hat G_{\rm{spin}}$ along $L_3$ of fig.  \ref{phase}. (b) sketch of the Coulomb phase.}
\label{Coulomb}
\end{figure}

The Gauss's law in the Coulomb phase of the monopole-spin hybrid system
reflects the fact that each monopole is neighbored with two spins, 
whereas each spin connects two monopoles of opposite
type, of which the spins are aligned by the singular magnetic field of the two
monopoles and point from the north to the south monopole. The Gauss's law is
then encapsulated in:
\begin{equation}
\begin{aligned}
\hat{G}_{\rm{mnp}}=\sum_{i=1}^{N}\Tilde{\sigma}_{z,i}\left(\sigma_{z}^{i+1}-\sigma_{z}^{i-1}\right),\\
\hat{G}_{\rm{spin}}=\sum_{i=1}^{N}{\sigma}_{z}^{i}\left(\Tilde{\sigma}_{z,i+1}-\Tilde{\sigma}_{z,i-1}\right).
\end{aligned}
\end{equation}
Where $\Tilde {\sigma}_{z,i}=D_i-H_i$.
We verify the existence of the Coulomb phase by determining the expectation value of
$\hat{G}_{\rm{mnp}}$ and $\hat{G}_{\rm{spin}}$ along the cutting line $\rm{L}_3$ in the phase diagram
(solid green line in Fig. \ref{phase}), which crosses the phases FM, CP and DW.
The results are
shown in Fig. \ref{Coulomb} (a). In the parameter regime out
of the CP the corresponding expectation values vanish and in the CP they approach unity. This verifies
that the Coulomb phase indeed pops up in the corresponding parameter regime,
and the spatial arrangement in the CP can be illustrated as in Fig. \ref{Coulomb} (b), which
is four-fold degenerate due to the periodic boundary condition applied. 

In the quantum simulation of lattice gauge field models with ultracold
atoms in the double-well superlattice 
\cite{Pan2020,Pan2022,Paulson2021,Halimeh2022,Zhai2022,Zhai2023},
the spatial configuration of the charges and photons in the CP is imprinted by 
an additional lattice and is actually an excited state of the system. 
In the DLG system, however, the CP is self-assembled by the DDI and is the true ground state.
The comparison of the CP realized in the double-well superlattice and the DLG system
demonstrates the flexibility of Hamiltonian engineering by the anisotropy effect of DDI.

\section{Zoo of quasiparticles in Coulomb phase}\label{zoo}
In this section, we demonstrate that the Coulomb phase hosts a zoo of
different quasiparticles, which can find potential application in
simulating various particle physics effects and phenomena.
Quasiparticles play an essential role in understanding the dynamical behavior of
quantum matter, ranging from superliquid  Helium \cite{Laudau1941,Feynman1954,Tucker1992, Adamenko2009}, condensed matter  \cite{Maiti2015,Hong2017,Peter2018,Rodin2020,Petar2023},
to ultracold atomic ensembles \cite{Alon2005,Choudhury2020,Singh2021,Van2023}. 
The concept of a quasiparticle is also closely connected to that of elementary particles \cite{Venema2016},
and this connection paves the way for simulating high energy physics with condensed matter 
and ultracold atomic systems. In this viewpoint, it is highly desirable to generate different types
of quasiparticles on a single platform, which can broaden the scope of the quantum simulation of high energy physics. 
The various quasiparticles arising in the Coulomb phase can find potential
applications in simulations of  e.g. vacuum fluctuations, 
 particle conversion and decay processes. 

\begin{figure*}[htbp]
\includegraphics[trim=65 350 55 10,width=0.95\textwidth]{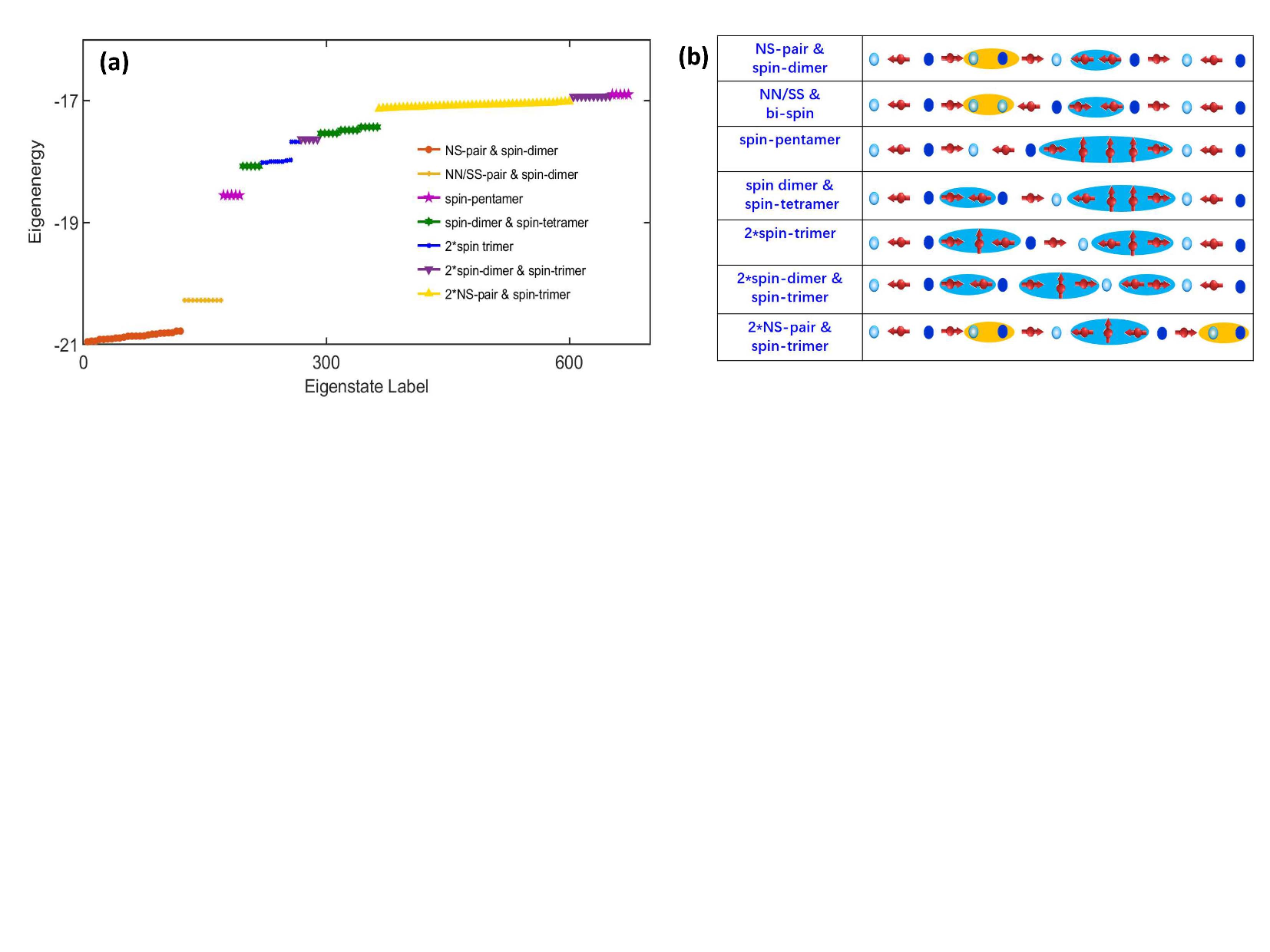} 
\caption{(color online). (a) The eigenenergy spectrum of the low-lying excited eigenstates.
The eigenstates corresponding to the same quasiparticle excitation are marked with the
same symbol. (b) Sketches of different quasiparticle excitations in the Coulomb phase. }
\label{spectrum}
\end{figure*}

The excitation of different quasiparticles can be captured in the eigenenergy
spectrum of the Coulomb phase, as shown in Fig. \ref{spectrum} (a),
in which a well-gaped band-like structure is presented. 
Each band of the spectrum corresponds to a particular type of quasiparticle excitation,
and in Fig. \ref{spectrum} (a) the eigenstates of the same quasiparticle
excitations are marked with corresponding colors.
The spectrum exhibits a rich detailed structure, see the band splitting
into subbands, such as the bands corresponding to the spin-pentamer excitation
(purple stars in Fig. \ref{spectrum} (a)) and the coexcitation of spin-dimer and 
spin-tetramer (green hexagram in Fig. \ref{spectrum} (a)), as well as the merging
of bands with different quasiparticle excitations,
e.g. the band of the pair excitation of spin-trimers (blue dots in Fig. \ref{spectrum} (a))
merging with that of
the co-excitation of spin-dimer and spin-tetramer.
The band merging indicates
the (quasi)degeneracy of different quasiparticle excitations, and can be 
explored to simulate particle conversion processes.
In the following we will address the quasiparticle excitations band-by-band
starting from the bottom of the spectrum.

The lowest excitation band in the spectrum corresponds to
the coexcitation of a monopole pair and a spin pair. The monopole pair is composed
of a north and a south monopole, and the two bounded monopoles behaves
as an emergent quasiparticle, which we term as the NS pair in the following. 
The pair of two spins is also manifested as a quasiparticle, 
and is referred to as the spin-dimer.
The second excited band is dominated by the co-excitation of the spin-dimer
and a bounded pair of monopole of the same type, composed of two
north or two south monopoles, which we term as the NN or SS pair, respectively.
The quasiparticles of the two lowest excitation bands are
sketched in the first two rows of Fig. \ref{spectrum} (b). 
It can be found that the local Gauss's law is maintained
in these two bands, with the spin neighbored to the north (south) monopole
pointing away from (to) the corresponding monopole, 
and the preservation of the local Gauss's law is attributed to
the singular magnetic field of the monopole.
Moreover, it can also be noticed that the spin-dimer resides to different
alignment in the two bands, either pointing to the same or the opposite directions,
which contributes an additional degree of freedom to the spin-dimer. 

The third to the sixth excited bands are domonated by the excitation of bounded
spin clusters of different sizes, of which the spatial configuration is
sketched in Fig. \ref{spectrum}(b).
The third excited band corresponds to the excitation of
a cluster of five spins, termed as spin-pentamer. 
The fourth band is then dominated by the coexcitation
of a spin-dimer and a cluster of four spins, dubbed as the spin-tetramer.
The excitation of the fourth band is through the conversion of a spin-pentamer
and a single spin to a spin-dimer and spin-tetramer, 
which conserves the total number of spins.
The spin-tetramer can further absorb a single spin and 
decay to two quasiparticles of spin-dimer and spin-trimer, 
i.e. a cluster of three spins, which forms the sixth excited band. 
The even higher bands contain multiple excitations of the aforementioned
quasiparticles, and for instance, the seventh excited band is dominated
by the excitation of two NS pairs with a spin-trimer.

\begin{figure*}[htbp]
\includegraphics[trim=35 140 50 0,width=0.95\textwidth]{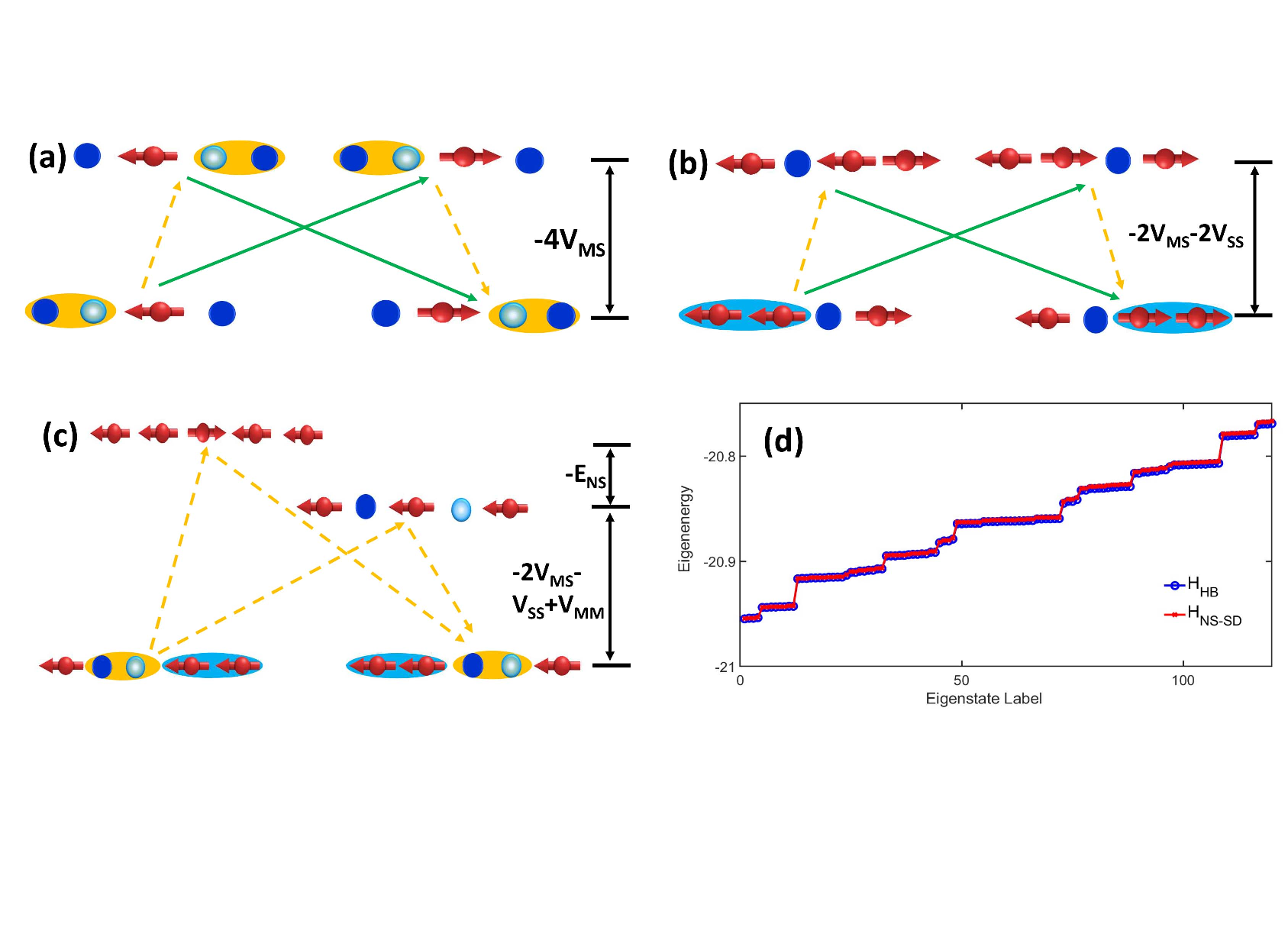}  
\caption{(color online). Channels for the second-order processes of NS pair hopping (a),
 the spin-dimer hopping (b) and the exchange interaction between the two quasiparticles (c).
 (d) Comparison between the energy spectrum of the first excited band obtained by 
 the complete Hamiltonian of the monopole-spin hybrid system and the derived effective Hamiltonian.}
\label{NS_bispin}
\end{figure*}

We take the NS-pair and the spin-dimer in the first excited band 
as an example to provide 
a microscopic description of the mobility and the mutual interaction
between these quasiparticles.
The NS pair and the spin-dimer both possess two degrees of freedom,
and the basis states are specified by two quantum numbers.
The basis state of the NS-pair is given as $|i,\sigma\rangle_{\rm{NS}}$, 
of which $i$ indicates the NS-pair located at the $i$ and $i+1$ sites of
the lattice  and $\sigma=\leftarrow(\rightarrow)$ denotes the north monopole
lying to the left (right) of the south monopole.
Similarly the basis state of the spin-dimer is defined as $|i,\sigma\rangle_{\rm{SD}}$,
which specifies the quasiparticle locating at $i$ and $i+1$ sites and 
both spins aligning to $\sigma$-direction, with $\sigma=\leftarrow/\rightarrow$. 

The hopping of the NS-pair is described by a second order process,
including a monopole exchanging position with its neighboring spin and then
flipping the direction of the spin by the  Dirac-string effect. This second-order
process is shown in Fig. \ref{NS_bispin}(a). Figure \ref{NS_bispin} (b) shows
the hopping of the spin-dimer, which turns out to be also a second-oder process,
with the spin hopping and flipping. 
It is interesting to notice that the hopping of both the NS-pair and the spin-dimer
represents a spin-orbit-coupling (SOC) like property, of which the hopping
of the quasiparticles is accompanied by the flipping of the inner state. Moreover,
the NS-pair and spin-dimer interact with each other by a second-order process, 
as sketched in Fig. \ref{NS_bispin}(c). 
The effective Hamiltonian describing the NS-pair and
spin-dimer can be derived as:
\begin{equation}
\begin{aligned}
\hat H_{\rm{NS-SD}}&=-J_{\rm{01}}\sum_{i=1}^{N}{\left(
\hat a^\dagger_{i}\hat a_{i+2}+\rm{H.c.}\right)\hat \sigma_{NS,x}
}\\
&-J_{\rm{02}}\sum_{i=1}^{N}{\left(
\hat b^\dagger_{i}\hat b_{i+2}+\rm{H.c.}\right)\hat \sigma_{SD,x}
}\\
&-J_{\rm{03}}\sum_{i=1}^{N}{\left(
\hat a^\dagger_{i}\hat a_{i+1}
\cdot\hat b^\dagger_{i}\hat b_{i-3}+H.c.\right)},\label{Hspindimer}
\end{aligned}
\end{equation}
where $\hat a^\dagger_{i}\hat a_{i+2}$ ($\hat b^\dagger_{i}\hat b_{i+2}$)
denotes the hopping of the NS pair (spin-dimer) from site-$i+2$ to site-$i$,
and $\hat \sigma_{NS,x}$ ($\hat \sigma_{SD,x}$) corresponds to the Pauli
matrix acting on the inner state of the NS pair and spin-dimer.
The hopping and interaction strengths can be derived as 
$J_{01}=2JJ_1/\left(-4V_{\rm{MS}}\right)$,
$J_{02}=2JJ_1/\left({-2V_{\rm{MS}}-2V_{\rm{SS}}}\right)$, 
$J_{03}=J_1^2/\left({-2V_{\rm{MS}}-V_{\rm{SS}}+V_{\rm{MM}}}\right)+{J_1^2}/\left({-2V_{\rm{MS}}-V_{\rm{SS}}+V_{\rm{MM}}-E_{\rm{NS}}}\right)$.
In Fig. \ref{NS_bispin}(d) we compare the eigenenergy spectrum of the first excited band obtained
from the original and the effective Hamiltonian, and the results agree very well, indicating that the effective Hamiltonian captures the
microscopic picture of the NS-pair and spin-dimer.

In this section, we have looked into the quasiparticle excitations corresponding to
the first seven excited bands of the eigenenergy spectrum of the Coulomb phase. These
excited bands present a quasiparticle zoo composed of bounded monopoles and 
bounded spin clusters, and these quasiparticles also dominate the intermediate higher
excited bands, in which the coexcitations of these quasiparticles take place.
It is worth mentioning that in the even higher energy regime of the spectrum 
new types of quasiparticles appear, which even break the local Gauss's law.
This indicates that the DLG-based monopole-spin hybrid system could simulate 
and test models beyond the current particle physics scope.

\section{Summary and discussion}\label{summary}
We have explored the dipolar ladder lattice gas for the quantum simulation of 
the monopole-spin hybrid system. In this scheme, both the spin and monopole excitations
are mapped to the spatial occupation degree of freedom, which allows to simulate the particle conversion
between the two types of particles by the hopping of ultracold atoms in the lattice,
as well as engineer the spin-spin, monopole-monopole and monopole-spin interactions by
the DDI. The anisotropy of the DDI further enables a tuning of the interaction strengths
in a wide range and gives rise to a rich phase diagram. 
This simulation scheme highlights the flexibility of the Hamiltonian engineering with the DDI.
Moreover, the DLG system also bears the experimental feasibility, and can by directly implemented on,
e.g. the Hubbard quantum simulator of magnetic erbium atoms loaded in optical lattices.

The simulated monopole-spin hybrid system presents a new type of self-assembled Coulomb phase, 
which fulfills the local Gauss's law. The Coulomb phase is generated by tuning the competition
 between the interactions and the monopole excitation energies through the DDI.
A zoo of quasiparticles arises in the excitation spectrum of the Coulomb phase, and the quasiparticles
can be formed by different components, either the bounded monopoles or the spin clusters, while the spin clusters can also be of different sizes. 
The quasiparticle zoo holds the potential of emulating various dynamical processes in different areas of physics.

The authors would like to acknowledge T.Shi and Y. Chang for inspiring discussions. 
This work was supported by the Key Researh and Development Program of China 
(Grants No.2022YFA1404102, No. 2022YFC3003802 and No. 2021YFB3900204 ).
This work is also supported by the Cluster of  Excellence 'Advanced Imaging of Matter' of the Deutsche Forschungsgemeinschaft (DFG) - EXC 2056 - project ID 390715994.

\bibliographystyle{apsrev4-2}

\end{document}